\documentclass[aps,prl,twocolumn,superscriptaddress,amssymb,showpacs]{revtex4}

\usepackage{graphicx}

\begin{document}


\title{Entangled self-phase locked states of light}



\author{H.~H.~Adamyan}
\email[]{adam@unicad.am}
\affiliation{Yerevan State University, A. Manookyan 1, 375049,
Yerevan, Armenia} \affiliation{Institute for Physical Research,
National Academy of Sciences,\\Ashtarak-2, 378410, Armenia}

\author{G.~Yu.~Kryuchkyan}
\email[]{gkryuchk@server.physdep.r.am}
\affiliation{Yerevan State University, A. Manookyan 1, 375049,
Yerevan, Armenia} \affiliation{Institute for Physical Research,
National Academy of Sciences,\\Ashtarak-2, 378410, Armenia}


\begin{abstract}
We explore in detail the possibility of generation of
continuous-variable (CV) entangled states of light fields with
well-localized phases. We show that such quantum states, called
entangled self-phase locked states, can be generated in
nondegenerate optical parametric oscillator (NOPO) based on a
type-II phase-matched down-conversion combined with polarization
mixer in a cavity. A quantum theory of this device, recently
realized in the experiment, is developed for both sub-threshold
and above-threshold operational regimes. We show that the system
providing relative phase coherence between two generated modes
also exhibits different types of quantum correlations between
photon numbers and phases of these modes. We quantify the
entanglement as two-mode squeezing and show that CV entanglement
is realized for both unitary, non-dissipative dynamics and for
dissipative NOPO in the entire range of pump field intensities.
\end{abstract}


\maketitle

\section{INTRODUCTION}

It is now believed that entanglement of quantum composite systems
with continuous degree of freedom is the basis of most
applications in the field of Quantum Information \cite{q}.
Interest in continuous variable (CV) entanglement is being
extensively excited by successful experiments on quantum
teleportation based on two-mode squeezed states \cite {levon2} as
well as the experiments dealing with entanglement in atomic
ensembles \cite {atom1}. Since then, remarkable theoretical and
experimental efforts have been devoted to generating and
quantifying CV entangled states.

In this paper we propose a novel type of CV entangled states of
light-field, which contain well-localized phases. They are
different from the well-known entangled Einstein-Podolsky-Rosen
(EPR) states generated in a nondegenerate optical amplifier
\cite{levon4,levon5}, which exhibit large phase fluctuations. We
believe that such entangled states can be generated in a
self-phase locked nondegenerate optical parametric oscillator
(NOPO), based on the type-II phase-matched down-conversion and
additional phase localizing mechanisms stipulated by the
intracavity waveplate. The motivations for this study are the
following:

For the first time the CV entangled states of light were studied
in \cite {levon4} and demonstrated experimentally in \cite{levon5}
for nondegenerate optical parametric amplifier (NOPA). Then a CV
entanglement source was built from two single-mode squeezed vacuum
states combined on a beamsplitter \cite {levon2}. It is well known
that each of the orthogonally polarized and frequency degenerate
fields generated by NOPO is a field of zero-mean values. The phase
sum of generated modes is fixed by the phase of the pump laser,
while their phase difference undergoes a phase diffusion process
\cite {levon4} stipulated by vacuum fluctuations. As a rule, the
NOPO phase diffusion noise is substantially greater than the shot
noise level, that limits the usage of NOPO in precision
phase-sensitive measurements. Various methods based on phase
locking mechanisms \cite{mason,fabr,murad,zond,kry} have been
proposed for reducing such phase diffusion. In the comparatively
simple scheme realized in the experiment \cite{mason}, self-phase
locking was achieved in NOPO by adding an intracavity quarter-wave
plate to provide polarization mixing between polarized signal and
idler fields. The evidence of self-phase locking was provided
there by the high level of phase coherence between the signal and
idler fields. Following this experiment, the semiclassical theory
of such NOPO was developed in \cite{fabr}. Recently, the schemes
of multiphoton parametric oscillators based on cascaded
down-conversion processes in $\chi ^{(2)}$ media placed inside the
same cavity and showing self-phase locking have been proposed
\cite{murad}. As was demonstrated in \cite{zond}, the system based
on combination of OPO and second harmonic generation also displays
self-phase locking. The formation of self-phase locking and its
connection with squeezing in the parametric four-wave mixing under
two laser fields has been demonstrated in \cite{kry}. An important
characteristic of self-phase locked devices concerns the phase
structure of generated subharmonics. Indeed, the formation of the
variety of distinct phase states under self-phase locked
conditions has been obtained in the mentioned Refs.
\cite{mason,fabr,murad,zond,kry}. It was recently noted that the
schemes involving phase locking are potentially useful for precise
interferometric measurements and optical frequency division
because they combine fine tuning capability and stability of type
- II phase matching with effective suppression of phase noise.
That is why we believe it will be interesting to consider phase
locked dynamics also from the perspective of Quantum Optics and,
in particular, from the standpoint of production of CV
entanglement.

A further motivation for such task is connected with the problem
of experimental generation of bright entangled light. So far, to
the best of our knowledge, there is no experimental demonstration
of CV entanglement above the threshold of NOPO. The progress in
experimental study of bright two-mode entangled state from cw
nondegenerate optical parametric amplifier has been made in
\cite{levon8}. The theoretical investigation of CV entangled light
in transition through the generation threshold of NOPO is given in
\cite{new}. One of the principal experimental difficulties in
advance toward a high-intensity level is the impossibility to
control the frequency degeneration of modes above the threshold.
We hope that the usage of phase locked NOPO may open a new
interesting possibility to avoid this difficulty.

In this paper we report what is believed to be the first
investigation of self-phase locked CV entangled states. We develop
the quantum theory of self-phase locked NOPO, with decoherence
included, in application to the generation of such entangled
states. This scheme is based on the combination of two processes,
namely, type-II parametric down-conversion and linear polarization
mixer with cavity-induced feedback. The parametric down-conversion
is a standard technique used to produce an entangled photon pairs
as well as CV two-mode squeezed states \cite{levon2}. The beam
splitter including polarization mixer are also considered as
experimentally accessible devices for production of entangled
light-fields \cite{tan}. Besides these, there have been some
studies of a beam splitter for various nonclassical input states,
including two-mode squeezing states \cite{kim}. It is obvious, and
also follows from the results of the mentioned papers \cite
{murad,zond,kry}, that the operational regimes of the combined
system with cavity-induced feedback and dissipation drastically
differ from those for pure processes. We show below that analogous
situation takes place in the investigation of quantum-statistical
properties of a combined system such as the self-phase locked
NOPO.

The paper is arranged as follows. In Section II we formulate the
model of combined NOPO based on the processes of two-photon
splitting and polarization mixing, and present a semiclassical
analysis of the system. Section III is devoted to the analysis of
quantum fluctuations of both modes within the framework of
linearization procedure around the stable steady-state. In Section
IV we investigate the CV entangling resources of self-phase locked
NOPO on the base of two-mode squeezing for both sub-threshold and
above-threshold operational regimes. We summarize our results in
Section V.

\section{MODEL OF SELF-PHASE LOCKED NOPO}

As an entangler we consider the combination of two processes in a
triply resonant cavity, namely, the type - II parametric
down-conversion in $\chi ^{(2)}$- medium and polarization mixing
between subharmonics in lossless symmetric quarter-wave plate. The
Hamiltonian describing intracavity interactions is
\begin{eqnarray}
H &=&i\hbar E\left( e^{i\left( \Phi _{L}-\omega
t\right)}a_{3}^{+}-e^{-i\left( \Phi _{L}-\omega t\right)
}a_{3}\right)\nonumber \\
&&+i\hbar k\left(e^{i\Phi _{k}}a_{3}a_{1}^{+}a_{2}^{+}-e^{-i\Phi
_{k}}a_{3}^{+}a_{1}a_{2}\right)\nonumber \\
&&+\hbar \chi \left( e^{i\Phi _{\chi} }a_{1}^{+}a_{2}+e^{-i\Phi
_{\chi} }a_{1}a_{2}^{+}\right) , \label{OriginalH}
\end{eqnarray}
where $a_{i}$ are the boson operators for the cavity modes
$\omega_{i}$. The mode $a_{3}$ at frequency $\omega $ is driven by
an external field with amplitude $E$ and phase $\Phi _{L}$, while
$a_{1}$and $a_{2}$ describe subharmonics of two orthogonal
polarizations at degenerate frequencies $\omega /2$. The constant
$ke^{i\Phi _{k}}$ determines the efficiency of the down-conversion
process. Linear coupling constant denoted as $\chi e^{i\Phi
_{\chi} }$describes the energy exchange between the modes and
besides, $\chi $ is determined by the amount of polarization
rotation due to the intracavity waveplate, $\Phi _{\chi}$
determines the phase difference between the transformed modes. We
take into account the detunings of subharmonics $\Delta _{i}$ and
the cavity damping rates $\gamma _{i}$ and consider the case of
high cavity losses for pump mode ($\gamma _{3}\gg \gamma $).
However, in our analysis we take into account the pump depletion
effects. The reduced density operator $\rho $ within the framework
of the rotating wave approximation and in the interaction picture
is governed by the master equation
\begin{eqnarray}
&&\frac{\partial \rho }{\partial t}=\frac{1}{i\hbar }\left[H_{int},\rho %
\right]+\sum_{i=1,2}\gamma _{i}\left( 2a_{i}\rho
a_{i}^{+}-a_{i}^{+}a_{i}\rho -\rho a_{i}^{+}a_{i}\right)\nonumber \\
&&+\frac{k^{2}}{\gamma _{3}}\left( 2a_{1}a_{2}\rho
a_{1}^{+}a_{2}^{+}-a_{1}^{+}a_{1}a_{2}^{+}a_{2}\rho -\rho
a_{1}^{+}a_{1}a_{2}^{+}a_{2}\right) ,  \label{MasterEq}
\end{eqnarray}
where
\begin{eqnarray}
H_{int}&=&\hbar \Delta _{1}a_{1}^{+}a_{1}+\hbar \Delta
_{2}a_{2}^{+}a_{2}\nonumber \\
&+&i\hbar \frac{kE}{\gamma _{3}}(a_{1}^{+}a_{2}^{+}-a_{1}a_{2})
+\hbar \chi(a_{1}^{+}a_{2}+a_{1}a_{2}^{+}). \label{TransformedH}
\end{eqnarray}
Let us also note that this equation is rewritten through the
transformed boson operators $a_{i}\rightarrow a_{i}\exp \left(
-i\Phi _{i}\right) $, with $\Phi _{i}$ being $\Phi _{3}=\Phi
_{L}$, $\Phi _{2}=\frac{1}{2}\left(\Phi _{L}+\Phi _{k}-\Phi _{\chi
}\right)$, $\Phi _{1}=\frac{1}{2}\left(\Phi _{L}+\Phi _{k}+\Phi
_{\chi }\right)$. That leads to cancellation of the phases on the
intermediate stages of calculations. As a result, the Hamiltonian
$H_{int}$ depends only on real-valued coupling constants. The
appearance of the last term in the master equation means that the
influence of the adiabatically eliminated fundamental mode is
reduced to an additional loss mechanism for the subharmonic modes.

In order to proceed further, we now consider the phase-space
symmetry properties of the system. It is easy to check that the
interaction Hamiltonian satisfies the commutation relation $\left[
H_{int},U\left(\pi \right)\right] =0$ with the operator
$U\left(\theta \right)=\exp \left[ i\theta \left(
a_{1}^{+}a_{1}-a_{2}^{+}a_{2}\right) \right] $. Moreover,
analogous symmetry $\left[ \rho \left( t\right) ,U\left( \pi
\right) \right] =0$ takes place for the density operator of the
system, which obeys the master equation (\ref{MasterEq}). Using
this symmetry we establish the following selection rules for the
normal-ordered moments of the mode operators:
\begin{equation}
\left\langle a_{1}^{+k}a_{2}^{+m}a_{1}^{l}a_{2}^{n}\right\rangle
=0, \label{ZeroMoments}
\end{equation}
if $k+l+m+n\neq 2j$, $j=0,\pm 1,\pm 2,...$. Another peculiarity
of\ NOPO with Hamiltonian (\ref{OriginalH}) is displayed in the
phase space of each of the subharmonic modes. The reduced density
operator for each of the modes is constructed from the density
operator $\rho $ by tracing over the other mode $\rho
_{1\left(2\right)}=Tr_{2\left(1\right)}\left(\rho \right)$.
Therefore, we find that $\left[ \rho _{1}\left( t\right)
,U_{1}\left( \pi \right) \right] =\left[ \rho _{2}\left( t\right)
,U_{2}\left( \pi \right) \right] =0$, where $U_{i}\left(\theta
\right)=\exp \left(i\theta
a_{i}^{+}a_{i}\right),\;\left(i=1,2\right)$ is the rotation
operator. It is easy to check using these equations that the
Wigner functions $W_{1}$ and $W_{2}$ of the modes have a two-fold
symmetry under the rotation of the phase-space by angle $\pi $
around its origin,
\begin{equation}
W_{i}\left( r,\theta \right) =W_{i}\left( r,\theta +\pi \right) ,
\label{WignerSymmetry}
\end{equation}
where $r$, $\theta $ are the polar coordinates of the complex
phase space. Note, that such symmetry relationships
(\ref{ZeroMoments}), (\ref {WignerSymmetry}) radically differ from
those taking place for the usual NOPO without the quarter-wave
plate. Indeed, in this case, (case of $\chi =0$ in Eqs.
(\ref{OriginalH})-(\ref{TransformedH})) the Wigner functions
$W_{i}$ are rotationally symmetric and the symmetry properties
(\ref{ZeroMoments}) read as $\left\langle
a_{1}^{+k}a_{1}^{l}a_{2}^{+m}a_{2}^{n}\right\rangle =0$, if
$k-l\neq m-n$.

We perform concrete calculations in the positive P-representation
\cite{drum} in the frame of stochastic equations for the complex
c-number variables $\alpha _{i}$ and $\beta _{i}$ corresponding to
the operators $a_{i}$ and $a_{i}^{+}$
\begin{eqnarray}
\frac{\partial \alpha _{1}}{\partial t}&=&-\left(\gamma
_{1}+i\Delta
_{1}\right)\alpha_{1}\nonumber \\
&&+\left(\varepsilon -\lambda \alpha _{1}\alpha _{2}\right)\beta
_{2}-i\chi \alpha _{2}+R_{1}, \label{a1StochEq}
\end{eqnarray}
\begin{eqnarray}
\frac{\partial \beta _{1}}{\partial t}&=&-\left( \gamma
_{1}-i\Delta _{1}\right) \beta _{1}\nonumber \\
&&+\left( \varepsilon -\lambda \beta _{1}\beta _{2}\right) \alpha
_{2}+i\chi \beta _{2}+R_{1.}^{+} \label{b1StochEq}
\end{eqnarray}
Here $\varepsilon =kE/\gamma _{3}$, $\lambda =k^{2}/\gamma _{3}$.
The equations for $\alpha _{2}$, $\beta _{2}$ are obtained from (\ref{a1StochEq}%
), (\ref{b1StochEq}) by exchanging the subscripts
(1)$\rightleftarrows $(2). $R_{1,2}$ are Gaussian noise terms with
zero means and the following nonzero correlators:
\begin{equation}
\left\langle R_{1}(t)R_{2}(t^{\prime })\right\rangle =\left(
\varepsilon -\lambda \alpha _{1}\alpha _{2}\right) \delta \left(
t-t^{\prime }\right) , \label{R1R2Correlator}
\end{equation}
\begin{equation}
\left\langle R_{1}^{+}(t)R_{2}^{+}(t^{\prime })\right\rangle
=\left( \varepsilon -\lambda \beta _{1}\beta _{2}\right) \delta
\left( t-t^{\prime }\right) .  \label{R1PlusR2PlusCorrelator}
\end{equation}
In this approach the stochastic amplitude $\alpha _{3\text{ }}$is given by $%
\alpha _{3}=\left( E-k\alpha _{1}\alpha _{2}\right) /\gamma _{3}$. So, Eqs.(%
\ref{a1StochEq})-(\ref{R1PlusR2PlusCorrelator}) involve the
depletion effect of the pump mode, which leads to the appearance
of the above-threshold operational regime.

First, we shall study the steady-state solution of the stochastic
equations in semiclassical treatment, ignoring the noise terms for
the mean
photon numbers $n_{j0}$ and phases $\varphi _{j0}$ of the modes ($%
n_{j}=\alpha _{j}\beta _{j}$, $\varphi _{j}=\frac{1}{2i}\ln \left(
\alpha _{j}/\beta _{j}\right) $). The mean photon numbers read as
\begin{equation}
n_{10}^{\pm }=\frac{1}{\lambda }\left( \frac{\Delta _{2}}{\Delta _{1}}%
\right) ^{1/2}\left[ \sqrt{\varepsilon ^{2}-\left( \varepsilon
_{cr}^{\pm }\right) ^{2}+\widetilde{\gamma
}^{2}}-\widetilde{\gamma }\right] , \label{Phot1ClassicGeneral}
\end{equation}
\begin{equation}
n_{20}^{\pm }=\frac{1}{\lambda }\left( \frac{\Delta _{1}}{\Delta
_{2}}\right) ^{1/2}\left[
\sqrt{\varepsilon ^{2}-\left( \varepsilon _{cr}^{\pm }\right) ^{2}+%
\widetilde{\gamma }^{2}}-\widetilde{\gamma }\right] ,
\label{Phot2ClassicGeneral}
\end{equation}
where
\begin{equation}
\widetilde{\gamma }=\frac{\gamma _{1}}{2}\left( \frac{\Delta
_{2}}{\Delta
_{1}}\right) ^{1/2}+_{{}}\frac{\gamma _{2}}{2}\left( \frac{\Delta _{1}}{%
\Delta _{2}}\right) ^{1/2}.  \label{GammaTilda}
\end{equation}
As we can see from (\ref{Phot1ClassicGeneral}), (\ref
{Phot2ClassicGeneral}), $n_{10}^{\pm }$ and $n_{20}^{\pm }$ are
real and positive for $\varepsilon $ exceeding two critical points
\begin{eqnarray}
\left(\varepsilon _{cr}^{\pm}\right)^{2}&=&\gamma _{1}\gamma
_{2}+\Delta _{1}\Delta
_{2}+\chi ^{2}\nonumber\\
&&\mp \sqrt{4\chi ^{2}\Delta _{1}\Delta _{2}-\left( \gamma
_{1}\Delta _{2}-\gamma _{2}\Delta _{1}\right) ^{2}},
\label{EpsilonCritical}
\end{eqnarray}
and, besides, the solutions $\ n_{i0}^{+}$ and $n_{i0}^{-},$
(i=1,2) correspond to two distinct values $\varepsilon _{cr}^{+}$
and $\varepsilon _{cr}^{-}$ accordingly. The steady-state values
of the phases corresponding to each of the critical points
$\varepsilon _{cr}^{-},$ $\varepsilon _{cr}^{+}$ are obtained as
\begin{eqnarray}
\sin ( \varphi _{20}^{+}&-&\varphi _{10}^{+} )
=\sin( \varphi_{20}^{-}-\varphi_{10}^{-} )=\nonumber\\
&=&\frac{1}{2\chi }\left[ \gamma _{1}\left( \frac{\Delta _{2}}{%
\Delta _{1}}\right) ^{1/2}-\gamma _{2}\left( \frac{\Delta _{1}}{\Delta _{2}}%
\right) ^{1/2}\right] ,  \label{SinPhaseDiff}
\end{eqnarray}
\begin{equation}
\cos (\varphi _{20}^{\pm }+\varphi _{10}^{\pm })=\frac{1}{\varepsilon }\sqrt{%
\varepsilon ^{2}-\left( \varepsilon _{cr}^{\pm }\right) ^{2}+\widetilde{%
\gamma }^{2}}.  \label{CosPhaseSum}
\end{equation}
\ It is easy to check that these solutions exist for both modes
only if the following relation holds
\begin{equation}
4\chi ^{2}\Delta _{1}\Delta _{2}>\left( \gamma _{1}\Delta
_{2}-\gamma _{2}\Delta _{1}\right) ^{2}.
\label{SolutionCondition}
\end{equation}
Let us note that the steady-state solutions (\ref{SinPhaseDiff}),
(\ref {CosPhaseSum}) completely determine the absolute phases of
the orthogonally polarized modes, which are hence self-locked,
unlike the ordinary NOPO. These results are in accordance with the
ones obtained in \cite {mason,fabr}, but for another configuration
of NOPO. In the scheme proposed in \cite{mason} only the signal
and idler modes are excited in the cavity, while the pump field is
a travelling wave. Nevertheless, in the adiabatic regime
considered here there is correspondence between both schemes.
Indeed, it is not difficult to check that the results (\ref
{Phot1ClassicGeneral}), (\ref {Phot2ClassicGeneral}) transform to
the corresponding results of the mentioned scheme \cite {fabr} by
replacing the parameter $\varepsilon $ with the corresponding pump
field amplitude.

We now turn to the standard linear stability analysis of these
solutions, assuming for simplicity, the perfect symmetry between
the modes, provided that the cavity decay rates and the detunings
do not depend on the polarization ( $\gamma _{1}=\gamma
_{2}=\gamma $, $\Delta _{1}=\Delta _{2}=\Delta $). The stability
of the system is governed by the matrices $F$ and $F_{+},$ $F_{-}$
describing the dynamics of small deviations $\delta \alpha
_{i\text{ }}$and $\delta \beta _{i}$ from the semiclassical
steady-state solutions (see, Sec.III). We reach stability if the
real parts of eigenvalues of these matrices are positive. This
analysis displays an evident dependence on the sign of the
detunings $\Delta _{1}=\Delta _{2}=\Delta $. For the positive
detuning, $\Delta >0$, only the steady-state solutions $%
n_{10}^{-}=n_{20}^{-} $ and $\varphi _{20}^{-}-\varphi _{10}^{-}$,
$\varphi _{20}^{-}+\varphi _{10}^{-}$ are stable, while for the
case of negative detuning the stability holds for the solutions
with the (+) superscript. As this analysis shows, for either sign
of detuning, the threshold is reached at $\varepsilon \geqslant $
$\varepsilon _{th}$, where
\begin{equation}
\varepsilon _{th}=\sqrt{\left( \chi -\left| \Delta \right| \right)
^{2}+\gamma ^{2}},  \label{EpsilonThreshold}
\end{equation}
and the steady-state stable solution for mean photon numbers can
be written in the general form as
\begin{equation}
n_{0}=n_{10}=n_{20}=\frac{1}{\lambda }\left[ \sqrt{\varepsilon
^{2}-\left( \chi -\left| \Delta \right| \right) ^{2}}-\gamma
\right] . \label{Phot12Simplified}
\end{equation}
The phases are found to be
\begin{equation}
\varphi _{10}=\varphi _{20}=-\frac{1}{2}Arc\sin
\frac{1}{\varepsilon }\left( \chi +\left| \Delta \right| \right)
+\pi k,  \label{PhasesNegDelta}
\end{equation}
for $\Delta >0$. For the opposite sign of the detuning, $\Delta
<0$, as we noted, the mean photon numbers are given by the same
Eq.(\ref {Phot12Simplified}), while the phases read as
\begin{eqnarray}
\varphi _{10} &=&\frac{1}{2}Arc\sin \frac{1}{\varepsilon }\left(
\chi +\left| \Delta \right| \right) +\pi \left(
k+\frac{1}{2}\right) ,
\label{PhasesPosDelta} \\
\varphi _{20} &=&\frac{1}{2}Arc\sin \frac{1}{\varepsilon }\left(
\chi +\left| \Delta \right| \right) +\pi \left(
k-\frac{1}{2}\right) ,  \nonumber
\end{eqnarray}
($k=0,1,2,..$). In the region $\varepsilon \leqslant $
$\varepsilon _{th}$ the stability condition is fulfilled only for
the zero amplitude steady-state solution $\alpha _{1}=\alpha
_{2}=\beta _{1}=\beta _{2}=0$. So, the set of above-threshold
stable solutions for both modes have two-fold symmetry in the
phase-space which was indeed expected from symmetry arguments
(\ref{WignerSymmetry}).

Let us now consider the output behavior of the system for the
special scheme of generation, when the couplings of in- and
out-fields occur at only one of the ring-cavity mirrors. Taking
into account that only the fundamental mode is coherently driven
by the external field with $\langle \alpha _{3}^{in}\rangle
=E/\gamma _{3}$, while the subharmonic modes are initially in the
vacuum state, we obtain for the mean photon numbers (in units of
photon number per unit time) $n_{3}^{in}=E^{2}/2\gamma
_{3},\;n_{i}^{out}=2\gamma _{i}n_{i0}\;\left(i=1,2\right)$ and
hence $n_{1}^{out}=n_{2}^{out}=2\gamma n_{0}$. Accordingly,
parametric oscillation can occur above the threshold pump power
$P_{th}=\hbar \omega ^{3} E_{th}^{2}/2\gamma _{3}$, where the
threshold value of the pump field is equal to $E_{th}=\frac{\gamma
_{3}}{k}\sqrt{\left(\chi -\left| \Delta \right| \right)^{2}+\gamma
^{2}}$.

We are now in a position to study quantum effects in self-phase
locked NOPO and will state the main results of the paper
concerning CV entanglement.

\section{ANALYSIS OF QUANTUM FLUCTUATIONS}

The aim of the present section is to study the quantum-statistical
properties of self-phase locked NOPO in linear treatment of
quantum fluctuations. Quantum analysis of the system using $P$-
representation is standard. A detailed description of the method
can be found in \cite{drum}. We assume that the quantum
fluctuations are sufficiently small so that Eqs.(\ref{a1StochEq}),
(\ref{b1StochEq}) can be linearized around the stable
semiclassical steady state $\alpha _{i}\left( t\right) =\alpha
_{i}^{0}+\delta \alpha _{i}\left( t\right) $, $\beta _{i}\left(
t\right) =\beta _{i}^{0}+\delta \beta _{i}\left( t\right) $. This
is appropriate for analyzing the quantum-statistical effects,
namely CV entanglement, for all operational regimes with the
exception of the vicinity of threshold, where the level of quantum
noise increases substantially. It should also be emphasized at
this stage that in the above-threshold regime of self-phase locked
NOPO the steady-state phases of each of the modes are well-defined
in contrast to what happens in the case of ordinary NOPO, where
phase diffusion takes place. According to this effect, the
difference between the phases, as well as each of the phases, can
not be defined in the above-threshold regime of generation of the
ordinary NOPO. On the whole, the well founded linearization
procedure can not be applied for this case. Nevertheless, the
linearization procedure and analysis of quantum fluctuations for
ordinary NOPO become possible due to the additional assumptions
about temporal behavior of the difference between the phases of
the generated modes \cite{levon10}.

We begin with consideration of below-threshold operational regime,
for $E<E_{th}$, where the equations linearized around the
zero-amplitude solution can be written in the following matrix
form
\begin{equation}
\frac{\partial }{\partial t}\delta \alpha ^{\mu }=-F_{\mu \nu
}\delta \alpha ^{\nu }+R^{\mu }\left( \alpha ,t\right) ,
\label{daStochEq}
\end{equation}
where $\mu =1,2,3,4$ and $\delta \alpha ^{\mu }=\left( \delta
\alpha _{1},\delta \alpha _{2},\delta \alpha _{3},\delta \alpha
_{4}\right) =\left( \delta \alpha _{1},\delta \alpha _{2},\delta
\beta _{1},\delta \beta _{2}\right) $, $R^{\mu }=\left(
R_{1},R_{2},R_{1}^{+},R_{2}^{+}\right) $. The 4$\times $4 matrix
$F_{\mu \nu }$ is written in the block form
\begin{equation}
F=\left(
\begin{array}{cc}
A, & B \\
B^{\ast }, & A^{\ast }
\end{array}
\right)  \label{FMatrix}
\end{equation}
with 2$\times $2 matrices
\begin{equation}
A=\left(
\begin{array}{cc}
\gamma +i\delta, & i\chi \\
i\chi, & \gamma +i\delta
\end{array}
\right) ,\;\;B=\varepsilon \left(
\begin{array}{cc}
0, & 1 \\
1, & 0
\end{array}
\right) .  \label{AandBMatrices}
\end{equation}
The noise correlators are determined as
\begin{equation}
\left\langle R^{\mu }\left( \alpha ,t\right) R^{\nu }\left( \alpha
,t^{\prime }\right) \right\rangle =D_{\mu \nu }\left( \alpha
\right) \delta \left( t-t^{\prime }\right)  \label{R0Correlators}
\end{equation}
with the following diffusion matrix
\begin{equation}
D=\left(
\begin{array}{cc}
B, & 0 \\
0, & B^{\ast }
\end{array}
\right) .  \label{DMatrix}
\end{equation}

First, we calculate the temporal correlation functions of the
fluctuation operators. The expectation values of interest can be
written as the integral
\begin{equation}
\left\langle \delta \alpha ^{\mu }\left( t\right) \delta \alpha
^{\nu }\left( t^{\prime }\right) \right\rangle =\int_{-\infty
}^{\min \left( t,t^{\prime }\right) }d\tau \left( e^{F\left(
t-\tau \right) }De^{F^{T}\left( t^{\prime }-\tau \right) }\right)
_{\mu \nu }, \label{dadaCorrelator}
\end{equation}
$F^{T}$ being the transposition of the matrix $F$. The integration
over $d\tau$ can be performed using the following useful formula
for operators $DF^{T}=FD$, obtained by straightforward
calculation. As a consequence, we arrive at the expression
$De^{F^{T}t}=e^{Ft}D$. Finally we obtain
\begin{equation}
\left\langle \delta \alpha ^{\mu }\left( t\right) \delta \alpha
^{\nu }\left( t^{\prime }\right) \right\rangle =-\frac{1}{2}\left(
F^{-1}e^{F\left| t-t^{\prime }\right| }D\right) _{\mu \nu },
\label{dadaCorrelatorSimplified}
\end{equation}
and hence for $t=t^{\prime }$%
\begin{equation}
\left\langle \delta \alpha ^{\mu }\left( t\right) \delta \alpha
^{\nu }\left( t\right) \right\rangle =-\frac{1}{2}\left(
F^{-1}D\right) _{\mu \nu }.  \label{dadaMean}
\end{equation}

This formula, however, is not very convenient for practical
calculations. Therefore, we rewrite the correlation functions of the quantum fluctuations $%
\delta \alpha _{i}$, $\delta \beta _{i}$ in a more simple form
through the two-dimensional column vectors $\delta \alpha =\left(
\delta \alpha _{1},\delta \alpha _{2}\right) ^{T}$, $\delta \beta
=\left( \delta \beta _{1},\delta \beta _{2}\right) ^{T}$. Upon
performing the calculation, we finally arrive at
\begin{widetext}
\begin{eqnarray}
\left\langle \delta \alpha \left( \delta \alpha \right)
^{T}\right\rangle &=&\frac{\varepsilon }{S^{4}-4\Delta ^{2}\chi
^{2}}\left[ \gamma \left(
\begin{array}{cc}
-2\chi \Delta, & S^{2} \\
S^{2}, & -2\chi \Delta
\end{array}
\right) -i\left(
\begin{array}{cc}
\chi \left(S^{2} -  2\Delta ^{2}\right), & \Delta \left(S^{2} -  2\chi ^{2}\right) \\
\Delta \left(S^{2} -  2\chi ^{2}\right), & \chi \left(S^{2} -
2\Delta ^{2}\right)
\end{array}
\right) \right],  \nonumber \\
  \nonumber \\
\left\langle \delta \alpha \left( \delta \beta \right) ^{T
}\right\rangle &=&\frac{\varepsilon ^{2}}{2\left(S^{4}-4\Delta
^{2}\chi ^{2}\right)}\left(
\begin{array}{cc}
S^{2}, & -2\chi \Delta \\
-2\chi \Delta, & S^{2}
\end{array}
\right) , \label{dadbMeanFinal}
\end{eqnarray}
\end{widetext}
where $S^{2}$ is introduced as
\begin{equation}
S^{2}=\gamma ^{2}+\chi ^{2}+\Delta ^{2}-\varepsilon ^{2}.
\label{Determinant}
\end{equation}
Note, that $S^{2}>0$ in the below-threshold regime.

As an application of these results the mean photon number in the
below-threshold regime can be calculated as follows,
\begin{equation}
n_{1}=n_{2}=\frac{\varepsilon ^{2}S^{2} }{2\left(S^{4}-4\Delta
^{2}\chi ^{2}\right) }. \label{PhotQBelow}
\end{equation}

Next we focus on the mode locked regime, for $E>E_{th}$,
considering the Eqs. (\ref{a1StochEq}), (\ref{b1StochEq}) in terms
of the fluctuations $\delta n_{i}\left( t\right) =n_{i}\left(
t\right) -n_{i0}$ and $\delta \varphi _{i}\left( t\right) =\varphi
_{i}\left( t\right) -\varphi _{i0}$ of photon number and phase
variables. In this regime the dynamics described by the linearized
equations of motion actually decouples into two independent
dynamics for two groups of combinations $\delta n_{\pm }=\delta
n_{2}\pm \delta n_{1}$, $\delta \varphi _{\pm }=\delta \varphi
_{2}\pm \delta \varphi _{1}$. In fact, one has
\begin{equation}
\frac{\partial }{\partial t}%
{\delta n_{+} \choose \delta \varphi _{+}} =-F_{+} {\delta n_{+}
\choose \delta \varphi _{+}} + {R_{n+} \choose R_{\varphi +}},
\label{dfiPlus_dnPlusStochEq}
\end{equation}
\begin{equation}
\frac{\partial }{\partial t} {\delta n_{-} \choose \delta \varphi
_{-}}=-F_{-}{\delta n_{-} \choose \delta \varphi _{-}} +{R_{n_{-}}
\choose R_{\varphi _{-}}}.  \label{dfiMinus_dnMinusStochEq}
\end{equation}
The drift, $F_{\pm }$, and the diffusion matrices, $\left\langle
R_{i}\left( t\right) R_{j}\left( t^{\prime }\right) \right\rangle
=D_{ij}^{(+)}\delta
\left( t-t^{\prime }\right) $, $(i,j=n_{+},\delta \varphi _{+})$; $%
\left\langle R_{n}\left( t\right) R_{m}\left( t^{\prime }\right)
\right\rangle =D_{nm}^{(-)}\delta \left( t-t^{\prime }\right)
,(n,m=n_{-},\varphi _{-})$ are respectively
\begin{equation}
\ F_{+}=\left(
\begin{array}{cc}
2\lambda n_{0}, & 4n_{0}\varepsilon \sin (\varphi _{20}+\varphi _{10}) \\
0, & 2\left( \gamma +\lambda n_{0}\right)
\end{array}
\right) ,  \label{F+Matrix}
\end{equation}
\begin{equation}
F_{-}=\left(
\begin{array}{cc}
2\gamma , & 4n_{0}\chi \sin n/\Delta \\
\Delta /n_{0}, & 0
\end{array}
\right) ,  \label{F-Matrix}
\end{equation}
\begin{equation}
D^{\left( +\right) }=\left(
\begin{array}{cc}
4n_{0}\gamma ,~ -2\varepsilon \sin (\varphi _{20}+\varphi _{10}) &  \\
-2\varepsilon \sin (\varphi _{20}+\varphi _{10}),~ -\gamma /n_{0}
&
\end{array}
\right) .  \label{D+-Matrix}
\end{equation}

Due to the decoupling between $\left( +\right) $ and $\left(
-\right) $ combinations of the modes, we conclude that the
following temporal correlation functions are equal to zero,
$\left\langle \delta \varphi _{+}\left(t\right)\delta \varphi
_{-}\left(t^{\prime }\right)\right\rangle =\left\langle \delta
n_{+}\left(t\right)\delta n_{-}\left(t^{\prime
}\right)\right\rangle=\left\langle \delta
n_{\pm}\left(t\right)\delta \varphi_{\mp}\left(t^{\prime
}\right)\right\rangle=0$, and hence $\left\langle \left( \delta
\varphi _{1}\right) ^{2}\right\rangle =\left\langle \left( \delta
\varphi _{2}\right) ^{2}\right\rangle $, $\left\langle \left(
\delta n_{1}\right) ^{2}\right\rangle =\left\langle \left( \delta
n_{2}\right) ^{2}\right\rangle $. The other correlation functions
can be calculated in the same way as described above for the
below-threshold regime. The temporal correlation functions are
derived as
\begin{equation}
\left\langle{\delta n_{\pm }\left( t\right)  \choose \delta
\varphi _{\pm }\left( t\right) }\left( \delta n_{\pm }\left(
t^{\prime }\right) ,\delta \varphi _{\pm }\left( t^{\prime
}\right) \right) \right\rangle =-\frac{1}{2}F_{\pm
}^{-1}e^{-F_{\pm }^{-1}\left| t-t^{\prime }\right| }D_{\pm },
\label{dn+-dfi+-Correlator}
\end{equation}
and hence
\begin{equation}
\left\langle{\delta n_{\pm } \choose \delta \varphi _{\pm }}\left(
\delta n_{\pm },\delta \varphi _{\pm }\right) \right\rangle
=-\frac{1}{2}F_{\pm }^{-1}D_{\pm }.  \label{dn+-dfi+-Mean}
\end{equation}

Performing the concrete calculations for each of the cases $\Delta >0$ and $%
\Delta <0$, we arrive at the following results
\begin{widetext}
\begin{eqnarray}
\left\langle{\delta n_{+} \choose \delta \varphi _{+}}\left(
\delta n_{+},\delta \varphi _{+}\right) \right\rangle
&=&\frac{1}{4\lambda n_{0}(\gamma +\lambda n_{0})}\left(
\begin{array}{cc}
4n_{0}\left[ \gamma (\gamma +\lambda n_{0})+(\chi -\mid \Delta
\mid )^{2}\right] , & -\lambda n_{0}(\chi -\mid \Delta \mid ) sign(\Delta ) \\
-2\lambda n_{0}(\chi -\mid \Delta \mid )sign(\Delta ), & -\lambda
\gamma
\end{array}
\right) ,  \label{dn+dfi+MeanFinal}\\
\nonumber\\
\left\langle {\delta n_{-} \choose \delta \varphi _{-}} \left(
\delta n_{-},\delta \varphi _{-}\right) \right\rangle
&=&\frac{1}{4\mid \Delta \mid \chi }\left(
\begin{array}{cc}
4n_{0}\chi (\chi -\mid \Delta \mid ), & 2\chi \gamma sign(\Delta ) \\
2\chi \gamma sign(\Delta ), & \frac{1}{n_{0}}(\gamma ^{2}-\mid
\Delta \mid (\chi -\mid \Delta \mid )
\end{array}
\right) .  \label{dn-dfi-MeanFinal}
\end{eqnarray}
\end{widetext}

We see that the considered system displays different types of
quantum correlations in terms of the stochastic variables, namely,
between photon-number sum and phase sum in the modes, as well as
between photon-number difference and phase difference in the
modes. These results radically differ from the ones taking place
for usual NOPO, where the correlation between photon-number
difference and phase sum in the modes is realized. The results
obtained indicate the possibilities to produce entanglement with
respect to the new types of quantum correlations.

\section{CV ENTANGLEMENT IN THE PRESENCE OF PHASE LOCKING}

Let us now turn our attention to quantum statistical effects and
the entanglement production for the case of perfect symmetry between the modes $%
(\gamma _{1}=\gamma _{2}=\gamma ,$\ $\Delta _{1}=\Delta
_{2}=\Delta )$. We note that unlike the two-mode squeezed vacuum
state, the state generated in the above-threshold regime of NOPO
is non-Gaussian, i.e. its Wigner function is non-Gaussian
\cite{cirak}. Recently, it has been demonsrated \cite{simon}, that
some systems involving beam splitters also generate non-Gaussian
states. The general consideration of this problem for self-phase
locked NOPO seems to be very complicated. However, the mentioned
results allow us to conclude that the state generated in
self-phase locked NOPO is most probably non-Gaussian. So far, the
inseparability problem for bipartite non-Gaussian state is far
from being understood. On the theoretical side, the necessary and
sufficient conditions for the separability of bipartite CV systems
have been fully developed only for Gaussian states, which are
completely characterized by their first and second moments. To
characterize the CV entanglement we address to both the
inseparability and strong EPR entanglement criteria \cite{epr}
which could be quantified by analyzing the variances of the
relevant distance $V_{-}=V\left( X_{1}-X_{2}\right) $ and the
total momentum $V_{+}=V\left( Y_{1}+Y_{2}\right)$ of the
quadrature amplitudes of two modes $X_{k}=\frac{1}{\sqrt{2}}\left[
a_{k}^{+}\exp \left( -i\theta _{k}\right) + a_{k}\exp \left(
i\theta _{k}\right) \right]$, $Y_{k}=\frac{i}{\sqrt{2}}\left[
a_{k}^{+}\exp \left( -i\theta _{k}\right) - a_{k}\exp \left(
i\theta _{k}\right) \right]$, $\left(k=1,2\right)$, where
$V(X)=\left\langle X^{2}\right\rangle -\left\langle X\right\rangle
^{2}$ is a denotation for the variance and $\theta _{k}$ is the
phase of local oscillator for the k-th mode. The two quadratures
$X_{k}$ and $Y_{k}$ are non commuting observables. The
inseparability criterion for the quantum state of two optical
modes reads as \cite{epr}
\begin{equation}
V=\frac{1}{2}(V_{+}+V_{-})<1,  \label{EntanglementCriteria}
\end{equation}
i.e. indicates that the sum of variances drops below the level of
vacuum fluctuations. Since the states of the system considered are
non-Gaussian, the criterion (\ref{EntanglementCriteria}) is only
sufficient for inseparability. The strong EPR entanglement
criterion is quantified by the product of variances as
$V_{+}V_{-}<\frac{1}{4}$. We remind that the sufficient condition
for inseparability (\ref{EntanglementCriteria}) in terms of the
product of variances reads as $V_{+}V_{-}<1$, i.e. is weaker than
the strong EPR condition.

In order to obtain the general expressions for variances, we first
write them in terms of the boson operators corresponding to the
Hamiltonian (\ref {OriginalH}). We perform the transformations
$a_{i}\rightarrow a_{i}\exp \left( i\Phi _{i}\right)$, which
restore the previous phase structure of the intracavity
interaction. Using also the symmetry relationships (\ref
{ZeroMoments}) we find quite generally the variances at some
arbitrary quadrature phase angles $\theta _{1}$, $\theta _{2}$ as
\begin{equation}
V_{\pm }=V\pm R\cos (\Delta \theta), \label{V+-ByVandR}
\end{equation}
where
\begin{eqnarray}
V &=&\frac{1}{2}(V_{+}+V_{-})=1+2n  \nonumber \\
&-2\mid &\left\langle a_{1}a_{2}\right\rangle \mid \cos (\Sigma
\theta+\Phi _{\arg }),~ \label{Vby_a}
\end{eqnarray}
\begin{equation}
R=2\mathop{\rm Re}(\left\langle a_{1}^{2}\right\rangle e^{i\Sigma
\theta })-2\mid \left\langle a_{1}^{+}a_{2}\right\rangle \mid ,
\label{RBy_a}
\end{equation}
\begin{equation}
\Delta \theta = \theta _{2}-\theta _{1}-\Phi_{\chi}, ~\Sigma
\theta = \theta _{1}+\theta _{2}+\Phi _{l}+\Phi _{k},
\label{DeltaSigmaTetta}
\end{equation}
and $n=\left\langle a_{1}^{+}a_{1}\right\rangle =\left\langle
a_{2}^{+}a_{2}\right\rangle$ is the mean photon number of the
modes, $\Phi _{\arg }=\arg \left\langle a_{1}a_{2}\right\rangle $.
So, in accordance with the formula (\ref{V+-ByVandR}), the
relative phase $\Phi_{\chi} $ between the transformed modes gives
the effect of the rotation of the quadrature amplitudes angle
$\theta _{2}-\theta _{1}$.

Obviously, the variances $V_{\pm }$ and hence the level of CV
entanglement depend on all parameters of the system including the
phases. The minimal possible level of $V$ is realized for an
appropriate selection of the phases $\theta _{i}$, namely for
$\theta _{1}+\theta _{2}=-\arg \left\langle
a_{1}a_{2}\right\rangle -$ $\Phi _{l}-\Phi _{k}$. Further, in most
cases we assume that this phase relationship takes place, but do
not introduce new denotations for $V$ and $V_{\pm}$ for the sake
of simplicity. In this case, ($\Sigma\theta=-\Phi _{\arg }$), in
correspondence with the formula (\ref {V+-ByVandR}), the variances
$V_{\pm}$ depend only on the difference between phases $\Delta
\theta$ and we arrive at
\begin{equation}
V=1+2(n-\mid \left\langle a_{1}a_{2}\right\rangle \mid ),
\label{VminBy_a}
\end{equation}
\begin{equation}
R=2\mathop{\rm Re} (\left\langle a_{1}^{2}\right\rangle e^{-i\Phi
_{_{\arg }}})-2\mid \left\langle a_{1}^{+}a_{2}\right\rangle \mid
. \label{RBy_aForMin}
\end{equation}

For the NOPO without additional polarization mixing $\left\langle
a_{1}^{+}a_{2}\right\rangle =\left\langle a_{1}^{2}\right\rangle
=\left\langle a_{2}^{2}\right\rangle =0$ and hence the case of the
symmetric variances $V_{+}=V_{-}$ is realized. The phase-locked
NOPO generally has non symmetric uncertainty region. However, the
variances $V_{+}$ and $V_{-}$ become equal for the special case of
$\theta _{2}-\theta _{1}-\Phi_{\chi} =\frac{ \pi }{2}$, when the
inseparability condition reads as $V<1$, $\left(
V=V_{-}=V_{+}\right) $. We note, that the relative phase
$\Phi_{\chi}$ plays an important role in specification of the
entanglement.

\subsection{Entanglement in the case of unitary time-evolution}

\begin{figure}
\includegraphics[angle=-90,width=0.48\textwidth]{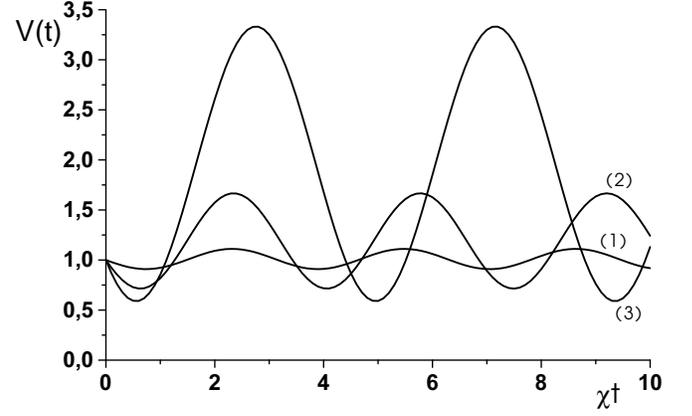}
\caption{ Unitary evolution of the variance $V\left(t\right)$
versus the scaled dimensionless interaction time $\chi t$ for the
range of $\varepsilon <\chi $ and provided that $\cos \left(
\Sigma\theta\right) =1$. The parameters are: $\varepsilon /\chi
=0.1$ (curve 1), $\varepsilon /\chi =0.4$\ (curve 2) and
$\varepsilon /\chi =0.7$ (curve 3). } \label{cos_fig}
\end{figure}

So far, we have considered mainly steady-state regime of
generation, including the effects of dissipation and
cavity-induced feedback. However, in order to better understand
the pecularities of the entanglement for the system under
consideration, it would be interesting and desirable to study the
case of a short interaction time $t<<\gamma ^{-1}$, when the
dissipation in the cavity is unessential. In our analysis we shall
assume that all processes are spontaneous, i.e. both subharmonic
modes are initially in the vacuum state. Then, the system's state
in the absence of losses is generated from the vacuum state by the
unitary transformation
\begin{equation}
\left| \Psi \left( t\right) \right\rangle =u\left( t\right) \left|
0\right\rangle _{1}\left| 0\right\rangle _{2}=e^{-\frac{i}{\hbar }%
H_{int}t}\left| 0\right\rangle _{1}\left| 0\right\rangle _{2},
\label{psi}
\end{equation}
which for $\chi \rightarrow 0$ gives two-mode squeezed vacuum
state. Using formula (\ref{Vby_a}), where $n=\langle
u^{-1}(t)a_{i}^{+}a_{i}u(t)\rangle $ and $\left\langle
a_{1}(t)a_{2}(t)\right\rangle =\langle
u^{-1}(t)a_{1}a_{2}u(t)\rangle ,$ and also considering the case of
zero-detunings $\Delta _{1}=\Delta _{2}=\Delta =0$,  after a long
but straightforward algebra we get the final result for the
variance $V=\frac{1}{2}(V_{+}+V_{-})$ for two different
operational regimes: $\varepsilon <$ $\chi $ and $\varepsilon >$
$\chi $. The variance for the range $\varepsilon <$ $\chi $ of a
comparatively weak scaled pump field $\varepsilon =kE/\gamma _{3}$
reads as
\begin{eqnarray}
V\left(t\right)&=&1-\frac{\varepsilon ^{2}}{\mu ^{2}}\left[\cos (2\mu t)-1\right]  \nonumber \\
&-&\frac{\varepsilon }{\mu }\sin (2\mu t)\cos \left(\Sigma\theta
\right) , \label{VShrodingerBelow}
\end{eqnarray}
where $\mu $ $=\sqrt{\chi ^{2}-\varepsilon ^{2}}$. Thus, we
observe a periodic evolution of $V$ in this regime typical for the
linear coupling. The level of squeezing of the two modes is
periodically repeated. The behavior of the two-mode variance also
significantly depends on the phase matching condition, so that we
may tune the phase sum to maximize the entanglement. We further
choose for illustrations the phase sum corresponding to $\cos
\left(\Sigma\theta\right) =1$ for both operational regimes. The
dependence of $V$ versus the scaled time-interval is shown in
Fig.~\ref{cos_fig}, where the three curves correspond to three
different choices of the ratio $\varepsilon /\chi $. Common to all
curves is that the variance is nonclassical and squeezed at least
at the points of its minima $t_{\min }$, which can be obtained by
the formula $ctg\left(2\mu t_{\min }\right)=\varepsilon /\mu $. In
all cases, the maximal degree of two-mode squeezing
$V_{\min}=V\left(t\right)|_{t=t_{\min}}=0.5$ is achieved in the
limit $\varepsilon \rightarrow \chi $. For the curves on the
Fig.~\ref{cos_fig} the time-intervals corresponding to the minimal
values of $V_{\min}$ equal to $\chi t_{\min
}\simeq{0.74}+{1.005}\pi k$(curve 1), $\chi t_{\min
}\simeq{0.63}+{1.091}\pi k$(curve 2), $\chi t_{\min
}\simeq{0.57}+{1.4}\pi k$(curve 3), ($k=0,1,2,...$).

If the opposite inequality holds, $\varepsilon >$ $\chi $, then
the nonlinear parametric interaction becomes dominant over the
linear coupling and the variance is given by the following formula
\begin{eqnarray}
V\left(t\right)&=&1+\frac{\varepsilon ^{2}}{\eta ^{2}}\left[\cosh \left(2\eta t\right)-1\right]  \nonumber \\
&-&\frac{\varepsilon }{\eta }\sinh\left(2\eta t\right)\cos \left(
\Sigma\theta\right) , \label{VShrodingerAbove}
\end{eqnarray}
where $\eta =\sqrt{\varepsilon ^{2}-\chi ^{2}}$.

\begin{figure}
\includegraphics[angle=-90,width=0.48\textwidth]{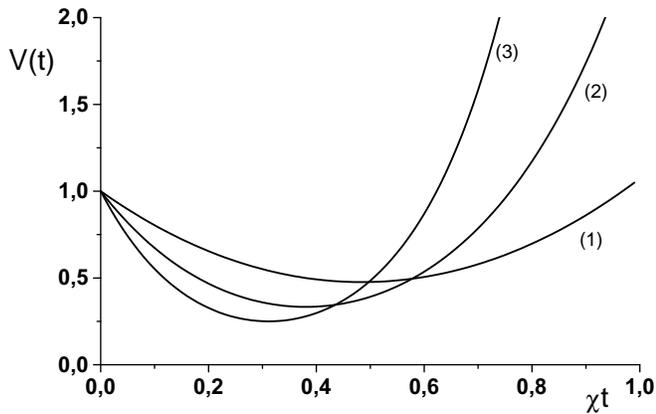}
\caption{ Unitary time evolution of the variance
$V\left(t\right)\;$ for the range of $\varepsilon >\chi$ and
provided that $\cos \left( \Sigma\theta\right) =1$. The parameters
are: $\varepsilon /\chi =1.1$ (curve 1), $\varepsilon /\chi =2$\
(curve 2) and $\varepsilon /\chi =3$ (curve 3).} \label{exp_fig}
\end{figure}

For $\chi \rightarrow 0$ and $\cos \left( \Sigma\theta\right) =1$,
from this formula we arrive at a well-known result,
$V\left(t\right)=\exp \left( -2\varepsilon t\right) $, for
two-mode squeezed state. This shows that in the limit of infinite
squeezing the corresponding state approaches to a simultaneous
eigenstate of $X_{1}-X_{2}$ and $Y_{1}+Y_{2}$ , and thus becomes
equivalent to the EPR state. Fig.~\ref{exp_fig} shows the behavior
of $V$ when the system operates in the regime $\varepsilon >$
$\chi $. This figure clearly shows that as the interaction time
increases, the variance decreases and reaches its minimum. Then
the squeeze variance exponentially increases with the growth of
the interaction time. For the data in Fig.~\ref{exp_fig} we obtain
that the time intervals for which the variance reaches the minima
are $\chi t_{\min }\simeq{0.48}$(curve 1), $\chi t_{\min
}\simeq{0.38}$ (curve 2), $\ \chi t_{\min }\simeq{0.31}$(curve 3).
It is easy to check that these points of minima can be found by
the formula $ctgh(2\eta t_{\min })=\varepsilon /\eta$, provided
that $\cos \left( \Sigma\theta\right) =1$.

We also conclude that the variance squeezes up to a certain
interaction time, if $\varepsilon >$ $\chi $. It should be noted
that although the time evolution of the variances are quite
different for each of the operational regimes, the minimal values
of the variance are described by the formula which is the same for
both regimes
\begin{equation}
V_{\min}=V\left( t \right)|_{t = t _ {\min} }=\frac{\chi
}{\varepsilon +\chi }. \label{VminShrodinger}
\end{equation}

With increasing $\varepsilon /\chi $, in the regime $\varepsilon
>$ $\chi $, the minimal value of the variance decreases as
$V_{\min}\sim $ $\chi /\varepsilon \ll 1$, which means that
perfect squeezing takes place in the limit of infinite pump field.
This result is not at all trivial for the system considered, even
in the absence of dissipation and cavity-induced feedback, because
the insertion of polarization mixer usually destroys the two-mode
squeezing produced by nondegenerate parametric down-conversion.
The reason is that two-mode squeezed vacuum state is a
supperposition of two-photon Fock states $\left| n\right\rangle
_{1}\left| n\right\rangle _{2}$ and the polarization mixer
destroys the Fock states having the same number of photons, i.e.,
$n_{1}=n_{2}=n$. The detailed analysis of this problem can be
found, for example, in \cite{kim}

\subsection{Sub-threshold regime}

Using formulas (\ref{dadbMeanFinal}), (\ref{VminBy_a}), after some
algebra, we obtain the minimal variance in the following form
\begin{eqnarray}
V=1+\frac{\varepsilon \left( \varepsilon S^{2}-\sqrt{\gamma ^{2}
S^{4}+\Delta ^{2}\left(S^{2} - 2\chi ^{2}\right)^{2}}
\right)}{S^{4}-4\Delta^{2}\chi^{2}}. \label{VminBelow}
\end{eqnarray}

\begin{figure}
\includegraphics[angle=-90,width=0.48\textwidth]{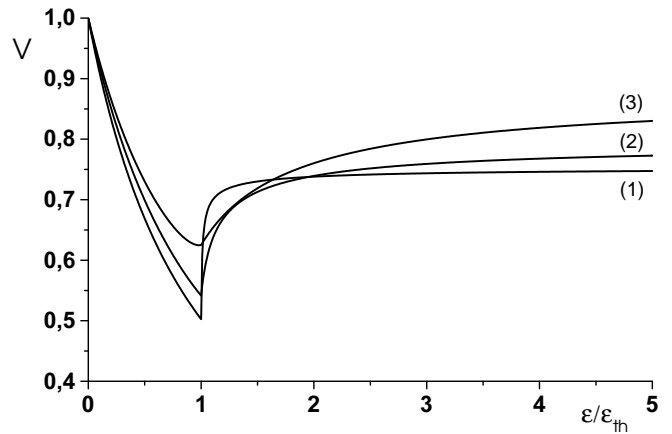}
\caption{Minimized variance $V$ versus dimensionless amplitude of
the pump field $\varepsilon /\varepsilon _{th}=kE/\gamma
_{3}\sqrt{(\chi -|\Delta |)^{2}+\gamma ^{2}}$ for both operational
regimes. The parameters are: $\chi /\gamma =0.1,\;\Delta /\gamma
=10\;$(curve 1), $\chi /\gamma =0.5,\;\Delta /\gamma =3$ (curve 2)
and $\chi /\gamma =0.5,\;\Delta /\gamma =1\;$(curve 3).}
\label{Vmin_fig}
\end{figure}

It is easy to check that in the limit $\chi \longrightarrow 0$\
the variance coincides with the analogous one for the ordinary
NOPO, $V=V_{-}=V_{+}=1-\varepsilon /(\varepsilon +\sqrt{\gamma
^{2}+\Delta ^{2}})$. We see that the minimal variance remains less
than unity for all values of pump intensity and is a monotonically
decreasing function of $\varepsilon ^{2}$. For all parameters the
maximal degree of two-mode squeezing $V\simeq 0,5$ is achieved
within the threshold range. It is also easy to check that this
expression is well-defined for all values of $\varepsilon ^{2}$,
including the vicinity of the threshold. One should keep in mind,
however, that the linear approach used does not describe the
threshold range where the level of quantum fluctuations increases.
As a consequence, some matrix elements of (\ref{dadbMeanFinal}),
(\ref{dn+dfi+MeanFinal}), (\ref{dn-dfi-MeanFinal}) increase
infinitely in the vicinity of threshold. Nevertheless, it follows
from (\ref{VminBelow}) and further results
(\ref{VminAbove})-(\ref{VMinusAbove}), that such infinite terms
are cancelled in the variance $V$, as well as in $V_{-},\;V_{+}$
for both operational regimes. This is not surprising, since such
cancellation of infinities in the quadrature amplitude variances
takes place for the ordinary NOPO also.

One of the differences between the squeezing effects of the
ordinary and self-phase locked NOPO is that the variances $V_{-}$
and $V_{+}$ for the ordinary NOPO are equal to each other, while
for the self-phase locked NOPO they are in general different. The
values of the non symmetric variances are expressed through $R$
according to formula (\ref{V+-ByVandR}). This quantity can be
calculated with the help of the formula (\ref{RBy_aForMin}). The
result is found to be

\begin{widetext}
\begin{equation}
R=\frac{\varepsilon \chi \Delta } {S^{4}-4\Delta^{2}\chi^{2}}
\left[ \frac { \varepsilon ^{4}-\left( \gamma ^{2}+\left(\Delta
-\chi \right)^{2}\right) \left( \gamma ^{2}+\left(\chi +\Delta
\right)^{2}\right) } { \sqrt{\gamma ^{2} S^{4}+\Delta
^{2}\left(S^{2} - 2\chi ^{2}\right)^{2} } } +2\varepsilon \right]
. \label{RForMinBelow}
\end{equation}
\end{widetext}

As we see, the final expressions (\ref{VminBelow}), (\ref
{RForMinBelow}) below the threshold are rather unwieldy. We show
the corresponding numerical results on the Figs.~\ref{Vmin_fig},
\ref{Vplusminus_fig}, \ref{VplusMulVminus_fig} for illustration.

\subsection{Above-threshold regime}

Performing calculations for each of the cases $\Delta >0$ and
$\Delta <0$, we find the variance $V$ in the above-threshold
regime in the following form
\begin{equation}
V=\frac{3}{4}-\frac{1}{4\sqrt{1+\left( \varepsilon
^{2}-\varepsilon _{th}^{2}\right) /\gamma ^{2}}}+\frac{\chi
}{4\mid \Delta \mid }. \label{VminAbove}
\end{equation}
The results (\ref{VminBelow}) and (\ref{VminAbove}) for both
operational regimes are summarized in Fig.~\ref{Vmin_fig}, where
the variance $V$ is plotted as a function of the amplitude of the
pump field. One can immediately grasp from the figure that the sum
of variances $V=\frac{1}{2}\left( V_{+}+V_{-}\right)$ remains less
than unity for all nonzero values of the pump field, provided that
$\chi< \left| \Delta \right|$. This shows the nonseparability of
the generated state. The maximal degree of entanglement is
achieved in the vicinity of threshold, $V\simeq 0,5$, if $\chi
/\left| \Delta \right| \ll 1$. In far above the threshold,
$E>>E_{th}$, $V$ increases with mean photon numbers of the modes
and reaches the asymptotic value $V=3/4+\chi /4\left| \Delta
\right|$. It should also be mentioned that the result
(\ref{VminAbove}) is expressed through the scaled pump field
amplitude $ \varepsilon =kE/\gamma _{3}$ and hence depends on
coupling constants $k $ and $\chi $.

\begin{figure}
\includegraphics[angle=-90,width=0.48\textwidth]{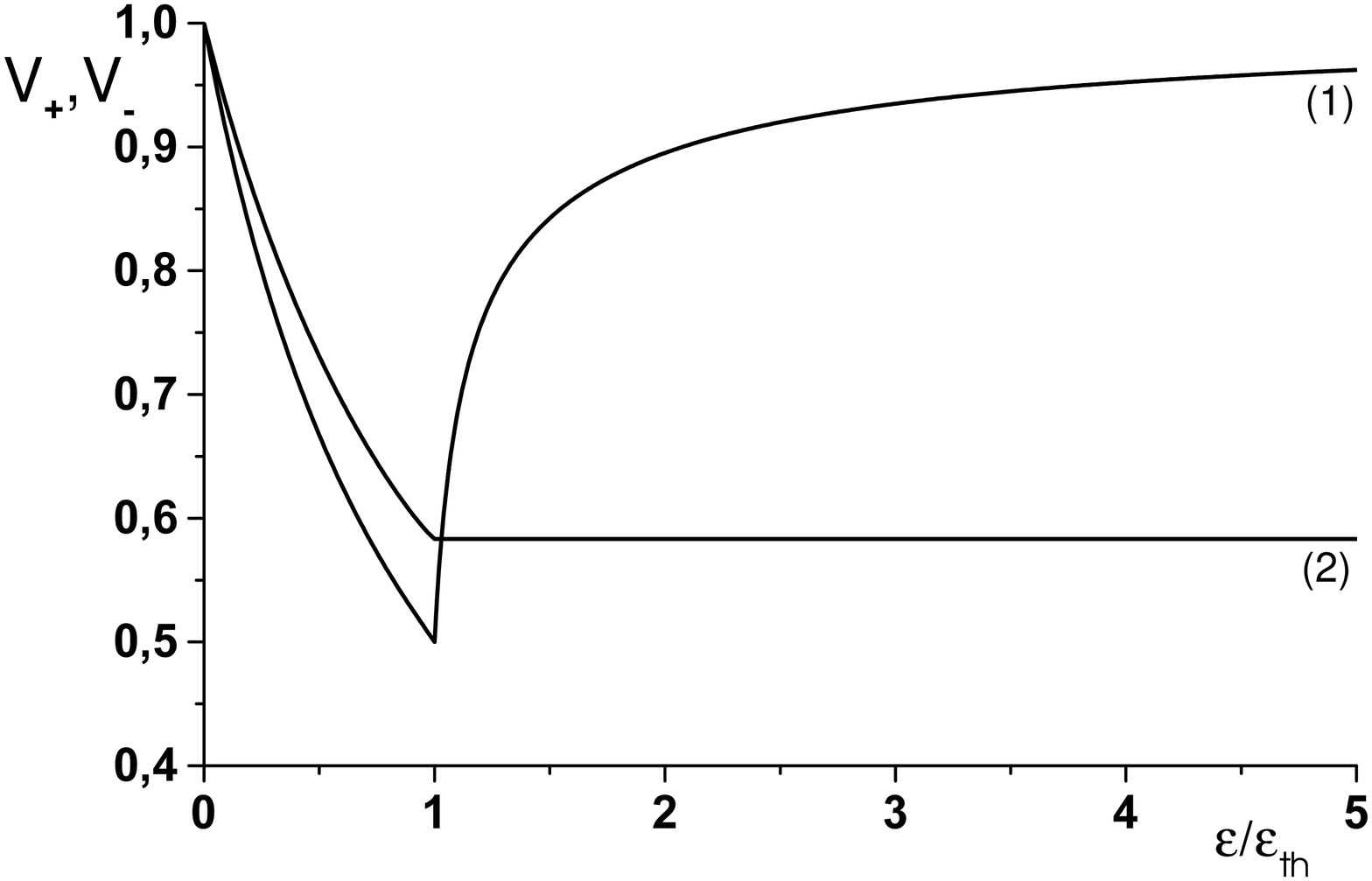}
\caption{ The variances $V_{+}\;$(curve 1) and $V_{-}\;$(curve 2)
versus the scaled pump field amplitude $\varepsilon /\varepsilon
_{th}\;$for the parameters: $\chi /\gamma =0.5,\;\Delta /\gamma
=3$.} \label{Vplusminus_fig}
\end{figure}

As the last step in connecting two-mode squeezing to observables
of CV entanglement, we report the expressions of the non symmetric
variances calculated with the help of the formulas
(\ref{V+-ByVandR}), (\ref{VminBy_a} ) and (\ref{RBy_aForMin}).
Upon evaluating all required expectation values, we obtain
\begin{equation}
R=\frac{sgn(\Delta )}{4}\left(\frac{\mid \Delta \mid - \chi }{\mid
\Delta \mid }-\frac{1}{\sqrt{1+\left( \varepsilon ^{2}-\varepsilon
_{th}^{2}\right) /\gamma ^{2}}}\right) \label{RForMinAbove}
\end{equation}
and hence
\begin{eqnarray}
V_{+} &=&\frac{3\mid \Delta \mid+\chi+ sgn\left(\Delta \right)\cos
\left(\Delta\theta\right)\left(\mid \Delta \mid-\chi\right)}{4\mid \Delta \mid} \nonumber \\
&-&\frac{{1}+sgn\left(\Delta \right)\cos
\left(\Delta\theta\right)}{{4}\sqrt{1+\left( \varepsilon
^{2}-\varepsilon _{th}^{2}\right) /\gamma ^{2}}} ,
\label{VPlusAbove}
\end{eqnarray}
\begin{eqnarray}
V_{-} &=&\frac{3\mid \Delta \mid+\chi- sgn\left(\Delta \right)\cos
\left(\Delta\theta\right)\left(\mid \Delta \mid-\chi\right)}{4\mid \Delta \mid} \nonumber \\
&-&\frac{{1}-sgn\left(\Delta \right)\cos
\left(\Delta\theta\right)}{{4}\sqrt{1+\left( \varepsilon
^{2}-\varepsilon _{th}^{2}\right) /\gamma ^{2}}} ,
\label{VMinusAbove}
\end{eqnarray}
For the case of $\Delta\theta=0$, when the variances are maximally
different, the results are reduced to
\begin{equation}
V_{-}=\frac{1}{2}+\frac{\chi }{2\Delta
},\;V_{+}=1-\frac{1}{2\sqrt{1+\left( \varepsilon ^{2}-\varepsilon
_{th}^{2}\right) /\gamma ^{2}}} \label{VPlusVMinusMaxDiff}
\end{equation}
for $\Delta <0$. The case of $\Delta >0$ is obtained from (\ref
{VPlusVMinusMaxDiff}) by exchanging $V_{+}\rightarrow V_{-}$\ and $%
V_{-}\rightarrow V_{+}$. These results for both operational
regimes are depicted on Fig.~\ref{Vplusminus_fig}. Some features
are immediately evident. First of all one can see that both
variances are minimal in the critical range but show quite
different dependences on the ratio $\varepsilon /\varepsilon
_{th}$ above the threshold. Attentive reader may ask about
dependence of the results obtained on parametric coupling constant
$k$. We note in this connection that the squeezed variances are
expressed through the scaled pump field amplitude $\varepsilon
=kE/\gamma _{3}$ and hence depend in general on both coupling
constants $k$ and $\chi $.

\begin{figure}
\includegraphics[angle=-90,width=0.48\textwidth]{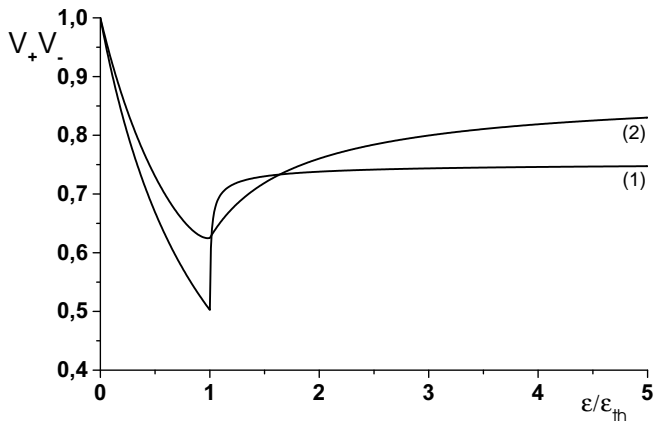}
\caption{ The product of the variances $V_{+}\;$ and $V_{-}\;$
versus the scaled pump field amplitude $\varepsilon /\varepsilon
_{th}\;$for the parameters $\chi /\gamma =0.1,\;\Delta /\gamma
=10\;$(curve 1) and $\chi /\gamma =0.5,\;\Delta /\gamma
=1\;$(curve 2).} \label{VplusMulVminus_fig}
\end{figure}

In terms of demonstrating the CV strong EPR entanglement, one has
to apply another criterion $V_{+}V_{-}<\frac{1}{4}$. For the case
of the symmetric uncertainties ($V_{+}=V_{-}=V$), the product of
the variances $V^{2}\geq \frac{1}{4}$ and hence the strong EPR
entanglement can not be realized. In the general case, the product
of the variances reads as
\begin{equation}
V_{+}V_{-}=V^{2}-R^{2}\cos ^{2}\left(\Delta\theta\right) .
\label{V+V-AboveGeneral}
\end{equation}
It seems that $V_{+}V_{-}$ lies below $1/4$ at least for the
relative phase $\Delta\theta=\pm \pi m$, $(m=1,2,...)$, and in the
vicinity of threshold, where $V\simeq 0,5$. However, for such
selection of the phases we arrive at
\begin{equation}
V_{+}V_{-}=\frac{\mid \Delta \mid + \chi}{4\mid \Delta \mid}
\left( 2- \frac{1}{\sqrt{1+\left( \varepsilon ^{2}-\varepsilon
_{th}^{2}\right) /\gamma ^{2}}}\right). \label{V+V-AboveFinal}
\end{equation}
It is easy to check that the product of the variances exceeds
$\frac{1}{4}$ even in the vicinity of the threshold. This quantity
for both operational regimes is illustrated in
Fig.~\ref{VplusMulVminus_fig}. It should be noted again that a
detailed analysis of this problem, must include more accurate
consideration of quantum fluctuations in the critical ranges.

\section{CONCLUSION}

Our work demonstrates the possibility of creation of CV entangled
light-states with well-localized phases. We show that such,
so-called, entangled self-phase locked states of light can be
generated in NOPO recently realized in the experiment
\cite{mason}. This device is based on the type-II phase-matched
down-conversion and additional phase localizing mechanisms
stipulated by the intracavity waveplate. The novelty is that this
device provides high level of phase coherence between the
subharmonics in contrast to what happens in the case of ordinary
NOPO, where the phase diffussion takes place. This development
paves the way towards the generation of bright CV entangled light
beams with well-localized phases. It looks like that this scheme
involving phase locking may be potentially useful for precise
interferometric measurements and quantum communications, because
it combines quantum entanglement and stability of type - II phase
matching with effective suppression of phase noise. The price one
has to pay for these advantages is the small aggravation of the
degree of CV entanglement in comparison with the case of ordinary
NOPO. The quantum theory of self-phase locked NOPO has been
developed in linear treatment of quantum fluctuations for both
below- and above-threshold regimes of generation. We have studied
the CV entanglement as two-mode squeezing and have shown that
entanglement is present in the entire range of pump intensities.
In all cases the maximal degree of two-mode squeezing $V\simeq
0,5$ is achieved in the vicinity of the threshold. It has also
been shown that the amount of the entanglement can be controlled
via the phase difference $\Phi_{\chi}$. The other peculiarities of
the system of interest have been established for the case of
unitary dynamics. One of these concerns the presence of two
operational regimes generating two-mode squeezing. If the linear
coupling between subharmonics dominates over the parametric
down-conversion, $\varepsilon<\chi$, we have observed a periodic
evolution of the squeezed variance. Maximal degree of squeezing
has been $V\simeq 0,5$ in this regime. If the parametric
interaction becomes dominant, $\varepsilon>\chi$, the more high
degree of two-mode squeezing can be obtained,
$V_{\min}=\frac{\chi}{\varepsilon+\chi}<{0.5}$, but up to certain
interaction time.

In our analysis we have not investigated all possible quantum
effects of self-phase locked parametric dynamics. In particular,
we have noted that the system considered displays different types
of quantum correlations, but we have not analyzed their connection
with all possible kinds of the entanglement. Consideration of
quantum fluctuations in the near threshold operational range of
self-phase locked NOPO also deserves special attention for more
accurate identification of CV entanglement. These topics are
currently being explored and will be the subject of forthcoming
work.

\begin{acknowledgments}
Acknowledgments: We thank J. Bergou for helpful discussions. This
work was supported by the NFSAT PH 098-02 grant no. 12052 and ISTC
grant no. A-823.
\end{acknowledgments}

\end{document}